\documentclass[aps,preprint]{revtex4}
\usepackage[dvipdf]{graphicx}

\begin{document}

\title{Order in the ground state of a simple cubic dipole lattice in an external field\footnote{This is the pre-peer reviewed version of an article that will appear in the International Journal of Quantum Chemistry. The manuscript was initially submitted under the title ``Ground state structure of a simple cubic dipole lattice in an external field''. This article may be used for non-commercial purposes in accordance with Wiley Terms and Conditions for Use of Self-Archived Versions.}}

\author{S. Ashhab}
\affiliation{Qatar Environment and Energy Research Institute, Hamad Bin Khalifa University, Qatar Foundation, Doha, Qatar}
\author{M. Carignano}
\affiliation{Qatar Environment and Energy Research Institute, Hamad Bin Khalifa University, Qatar Foundation, Doha, Qatar}
\author{M. E. Madjet}
\affiliation{Qatar Environment and Energy Research Institute, Hamad Bin Khalifa University, Qatar Foundation, Doha, Qatar}

\date{\today}

\begin{abstract}
Motivated by the presence of a lattice of rotating molecular dipoles in the high temperature phase of methylammonium lead iodide, we investigate the ground state of a simple cubic lattice of dipoles interacting with each other via the dipole-dipole interaction and with an external field via the Zeeman interaction. In the absence of an external field, the ground state is infinitely degenerate, and all the configurations in the ground state manifold are periodic along the three lattice axes with period 2. We numerically determine the ground state of a 1000-dipole lattice interacting with an external field, and we analyze the polarization, dipole orientation statistics and correlations in this state. These calculations show that for some special directions of the external field the two-site periodicity in the dipole configurations is preserved, while in the general case this periodicity is lost and complex dipole configurations form in the presence of the external field. 
\end{abstract}

\maketitle

\section{Introduction}
\label{Sec:Introduction}

Spin models are ubiquitous in theoretical physics in the study of a variety of physical systems, and they can be used to understand a wide range of phenomena such as ordered states of matter, domain boundaries and phase transitions \cite{SpinModelBooks}. Recently, they have also served as natural models for the study of entanglement and related properties in the study of condensed matter systems from the point of view of quantum information \cite{Amico,Eisert}. In the field of light harvesting devices, the recently emerging hybrid organic-inorganic lead-halide perovskites have opened a novel application for spin models because the organic component often has a net dipole moment. The most representative perovskite material in this area is CH$_3$NH$_3$PbI$_3$, which in its high temperature phase contains a simple cubic lattice of dipoles from the polar molecule CH$_3$NH$_3$ that occupies the A site in the perovskite crystal structure \cite{Baikie2013aa,Stoumpos2013aa,Ashhab}. Spin models have been used to explore the possible origin of hysteresis observed in solar cells \cite{Frost2014ab,Frost2014aa}, to understand the dynamics of the cation rotations \cite{Leguy} and also to explain enhancement of carriers conductivity at the interface between domains \cite{Rashkeev}. More recently, spin-models have been used to explore the long-range order in perovskites \cite{Lahnsteiner} and the structure of the interface between domains with different ordering patterns across the interface \cite{Ashhab}.

Spin models apply more generally to other systems in nature, some examples being the ordering of the electric dipoles of water molecules in ice \cite{Bramwell}, the dipolar ordering of small molecules trapped in regular cage structures \cite{Cioslowski,Aoyagi,Gorshunov}. Moreover, the application of spin models to dielectric and magnetic properties of materials has a long history \cite{Honda,Ornstein,Luttinger,Lax,Berlin,Nijboer,Horvat}. Artificially fabricated magnetic superstructures can also be described by similar models \cite{Skomski}. One particularly interesting result, found by Luttinger and Tisza \cite{Luttinger}, is that in the absence of an external field a simple cubic lattice of dipoles has an infinitely degenerate ground state with no net polarization. There have been several studies on the polarization properties of the ground state of a dipole lattice under the influence of an external field \cite{Luttinger,Lax,Berlin,Nijboer,Kukhtin,AbuLabdeh,Galkin,Bondarenko,Johnston}. These studies have generally considered cases where the external field points in some special directions that allow the derivation of analytical results. In the general case of an arbitrary field direction, one cannot perform fully analytical derivations to our knowledge, and numerical methods are necessary. Here we perform such numerical calculations on a dipole lattice in an external field, analyzing several different field directions. By evaluating dipole orientation statistics and correlation functions, we show that for a field direction outside any of the main lattice planes (i.e.~xy, yz and xz) of a simple cubic lattice the simple structure of the ground state is lost, in addition to the lifting of the degeneracy that exists at zero external field.

We emphasize from the outset that an accurate model of CH$_3$NH$_3$PbI$_3$ and similar materials would include the inorganic lattice and its interactions with the organic molecular dipoles. However, our purpose in this work is to gain an understanding and insight into the physical mechanisms related to the dipole-dipole interactions in the system. The results that we obtain in this work can then be used to contribute to our understanding of the role that dipolar interactions could play in such a complex material.

\section{Ground state in the absence of an external field}
\label{Sec:ZeroField}

We consider a simple cubic lattice of dipoles interacting via the dipole-dipole interaction. For definiteness, we shall use the language of electric dipoles, although the results are general and apply e.g.~to magnetic dipoles. The total dipole-dipole interaction energy of the lattice is given by
\begin{equation}
U_{\rm int} = \frac{1}{8\pi\epsilon} \sum_{i,j} \frac{\vec{p}_i\cdot\vec{p}_j - 3 \left(\vec{p}_i\cdot\hat{r}_{ij}\right) \left(\vec{p}_j\cdot\hat{r}_{ij}\right)}{\left|r_{ij}\right|^3},
\label{Eq:DipoleDipoleInteractionEnergy}
\end{equation}
where $\epsilon$ is the permitivity, $\vec{p}_i$ is the dipole vector of dipole $i$, $\hat{r}_{ij}$ is the unit vector pointing from the location of dipole $i$ to that of dipole $j$, and $r_{ij}$ is the distance between dipoles $i$ and $j$. In order to transform the Hamiltonian into a more universal form, we define the new variables
\begin{eqnarray}
\vec{p}_i & = & p \hat{p}_i \nonumber \\
r_{ij} & = & r_0 \tilde{r}_{ij} \\
U_0 & = & \frac{p^2}{4\pi \epsilon r_0^3}, \nonumber
\end{eqnarray}
where $p$ is the magnitude of the dipole moments (which is assumed to be fixed and equal for all the dipoles), and $r_0$ is the lattice parameter. The vector $\hat{p}_i$ is now the unit vector in the direction of dipole $i$, and $\tilde{r}_{ij}$ is the renormalized distance between dipoles $i$ and $j$ in units of the lattice parameter. We emphasize that the problem under study is a classical one, and the hat symbols should not be confused as indicating quantum mechanical operators. With these definitions the dipole-dipole total interaction energy is given by
\begin{equation}
U_{\rm int} = \frac{U_0}{2} \sum_{i,j} \frac{\hat{p}_i\cdot\hat{p}_j - 3 \left(\hat{p}_i\cdot\hat{r}_{ij}\right) \left(\hat{p}_j\cdot\hat{r}_{ij}\right)}{\left|\tilde{r}_{ij}\right|^3}.
\label{Eq:DipoleDipoleInteractionEnergyRenormalized}
\end{equation}
We keep the factor of 2 in the coefficient on the right-hand side to indicate that this factor is used to avoid double-counting the interaction energy of a single dipole pair.

The ground state of the above-described dipole lattice is infinitely degenerate \cite{Luttinger}. If we choose one dipole in the lattice and designate it as the origin of our system of coordinates, this dipole can point in any direction on the unit sphere with Cartesian components $p_x$, $p_y$ and $p_z$. The orientations of all other dipoles are then given by the following simple rule: the dipole at location $(x, y, z)$ [each of these being an integer] has Cartesian components $(-1)^{y+z} p_x$, $(-1)^{x+z} p_y$ and $(-1)^{x+y} p_z$. This dipole configuration has periodicity 2 along each of the crystal axes (i.e.~x, y and z). The periodicity in any of these three directions becomes 1 in the special case where the dipoles are oriented along that direction.

\begin{figure}[h]
\includegraphics[width=12.0cm]{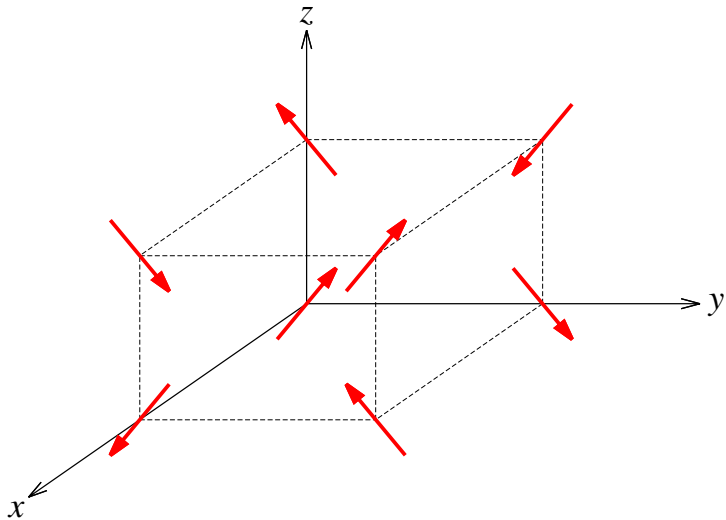}
\caption{Schematic diagram of the LT ground state. In the absence of an external field, one has a two dimensional manifold of degenerate ground states. The dipole at the origin can be chosen to point in any direction, which then determines the directions of all the dipoles in the lattice. The LT ground state has two-site translation symmetry in the x, y and z directions. The net polarization is zero in the absence of an external field.}
\label{Fig:LuttingerTizsaConfiguration}
\end{figure}

Figure \ref{Fig:LuttingerTizsaConfiguration} illustrates one example of the manifold of degenerate Luttinger-Tisza (LT) ground states, all of which possess the same energy ($-2.79 N U_0$ for an interaction cutoff of $3r_0$, with $N$ being the total number of dipoles) and no net polarization.

\section{Polarization and dipole configurations in the presence of an external field}
\label{Sec:FiniteField}

We now calculate and analyze the ground state of the dipole lattice under the influence of an externally applied field. 

The energy arising from the interaction between the dipole lattice and the external field is given by
\begin{equation}
U_{\rm ext} = - \vec{E} \cdot \sum_{i} \hat{p}_i,
\label{Eq:DipoleFieldInteractionEnergy}
\end{equation}
where the vector $\vec{E}$ is the external field, including any factors that convert the field into the appropriate energy units. The total energy is now given by $U_{\rm total} = U_{\rm int} + U_{\rm ext}$.

Before analyzing the case of a finite external field, we mention that in the absence of the external field we can take the energy per dipole, i.e.~$-2.79 U_0$, and infer from it the value of the internal field induced by all neighbouring dipoles at the location of any given dipole. Since all the dipoles in the lattice are equivalent to each other in any of the LT configurations, all of the dipoles will have the same energies and experience the same absolute value of the local field. Using the formula $U_{\rm int} = - \vec{E}_{i, \rm int} \cdot \hat{p}_i \times N/2$, where the unspecified index $i$ indicates that we can choose any dipole in the lattice, we find that the internal field is $5.589 U_0$ pointing in the same direction as the dipole (which is true for every dipole in the lattice). Note that we again use the factor of 2 here as in Eq.~(\ref{Eq:DipoleDipoleInteractionEnergyRenormalized}), because the interaction energy of a pair is calculated as the energy of only one of the two dipoles in the field induced by the other dipole.

In our calculations, we generally use a $10 \times 10 \times 10$ lattice of dipoles. In each calculation we search for the dipole configuration that minimizes the energy for a given field direction and strength. We do so by initializing the dipole configuration in a state of randomly oriented dipoles and allowing the dipoles to relax to lower energy states: in each iteration of the calculation we evaluate the total field at each dipole location and then rotate each dipoles in the direction that maximizes the reduction in its energy. The rotation angle for each dipole is proportional to the torque felt by the dipole multiplied by an overall step size. This step size is reduced whenever the above procedure stops reducing the total energy, and the calculation is stopped when the step size reaches $10^{-6}$, at which point we expect that we have reached a good approximation of the ground state. We find that, with some exception explained below, we generally obtain the same dipole configuration for all randomly generated initial states. After obtaining a given optimized dipole configuration, we analyze its polarization and statistical properties.

\begin{figure}[h]
\includegraphics[width=12.0cm]{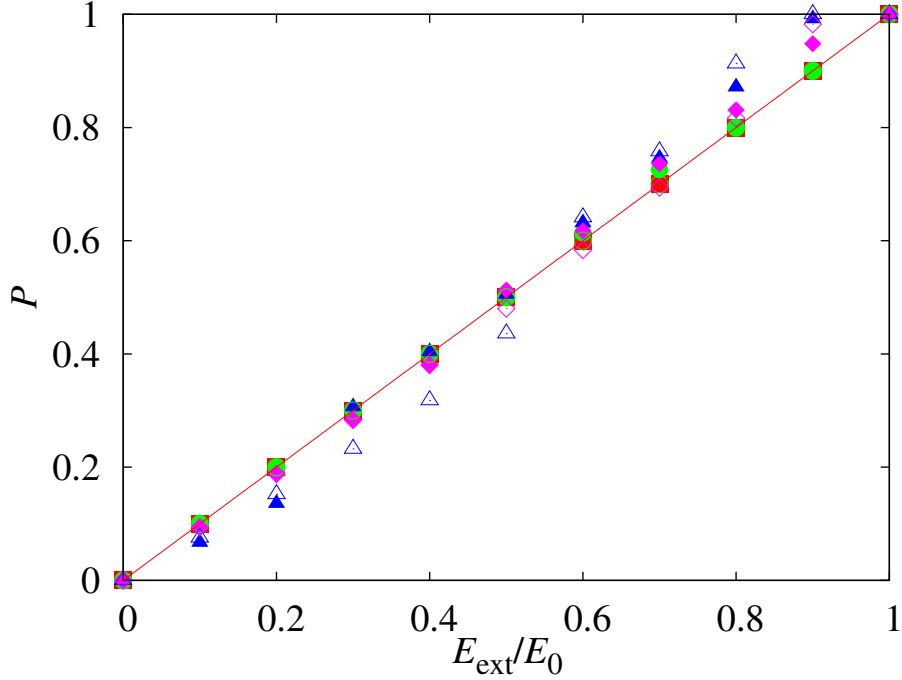}
\caption{The polarization $P$ as a function of external field strength for different field directions: along one of the main crystal axes [i.e.~x, y or z] (red squares), in one of the main planes and making an angle $\pi/8$ with one of the main axes (green circles), along the (1,1,1) direction (blue triangles) and in the direction $(\theta,\phi)=(\pi/6,\pi/6)$ (magenta diamonds). The filled symbols are obtained using a $10 \times 10 \times 10$ lattice, while the open symbols are obtained using a $2 \times 2 \times 2$ lattice. The solid red line is a straight line with slope equal to one. The red squares and green circles lie almost exactly on top of the line.}
\label{Fig:Polarization}
\end{figure}

The polarization, defined as $P=\left|\sum_i\hat{p}_i\right|/N$, as a function of external field strength is shown in Fig.~\ref{Fig:Polarization}. When the external field lies in one of the three planes xy, xz and yz (to which we refer as the main planes below), the polarization has a linear dependence on the field from $E=0$ to $E=E_0$, where $E_0$ is the value of the externally applied field at which the system reaches a state of full polarization along the direction of the applied field. When we take different field directions, i.e.~directions where the external field has finite components along x, y and z, we find that the polarization does not deviate much from the linear dependence, as also shown in Fig.~\ref{Fig:Polarization}. We note here that although this result agrees with intuitive expectations for dielectric and paramagnetic materials, to our knowledge it has not been shown to be true for a LT dipole lattice before. For all field directions, the polarization is parallel to the direction of the external field, except for small deviations at intermediate values of the external field.

It is interesting that the value of the external field $E_0$ at which the system becomes fully polarized is given by $5.589 U_0$, which is exactly the value of the internal field induced by neighbouring dipoles at zero field, even though these two fields occur in different situations. In other words, the dipole-induced field is given by $5.589 U_0$ only when the external field strength is zero; the dipole-induced field then decreases and reaches the value zero as the external field is increased to the value $5.589 U_0$. Based on this observation, one might wonder whether the total field at any dipole location remains constant between the two above limits. We have confirmed numerically that this is not the case. The local field decreases from $5.589 U_0$ to a smaller value as the external field strength is increased from zero, and then above a certain value of the external field the local field reverses its tendency and starts increasing to come back to the value $5.589 U_0$ when the external field reaches the value $5.589 U_0$ (meaning that the dipole-induced field vanishes at that point, as discussed in more detail in the Appendix).

In the following three subsections, we analyze the dipole configurations in some more detail.

\subsection{Weak external field}
\label{Sec:WeakField}

Let us take the infinitely degenerate ground states in the case of zero field discussed in Sec.~\ref{Sec:ZeroField}, i.e.~the LT state, and consider the effect of an infinitesimally weak field.

We start with the case where the externally applied field is parallel to one of the three lattice axes, and for definiteness we take this axis to be the z axis. In physical systems that have a degenerate ground state, a weak external field generally plays the role of a perturbation that lifts the degeneracy. We therefore ask: what dipole configuration does this external field favour? If we had a single dipole, which can point in any direction in the absence of an external field, the dipole would align with the external field. Somewhat counter-intuitively, in the case of the dipole lattice under study, the external field pointing in the z direction favours dipole configurations in which the dipoles point perpendicular to the z axis \cite{Luttinger,Kukhtin}. This result can be understood by considering that any given dipole feels an internal field that is induced by the neighbouring dipoles, and this field is parallel to the direction of the dipole under consideration. The weak external field acts as a small perturbation that slightly modifies the field felt by each dipole. In order for the infinitesimally weak external field to make the largest possible change in the total energy, it is favourable to have the external field perpendicular to the internal field. In this case, each dipole rotates by a small amount towards the direction of the external field, leading to an energy reduction relative to the unperturbed ground state. This situation can now be contrasted with that of a configuration in which the dipole orientations are all parallel to the z axis and we add a weak external field along the z axis. In this case, since the internal field points along the z axis, the external field slightly modifies the value of the field at each dipole location, but it does not change the direction of the field. As a result, each dipole will still point along the total field at its location, making each dipole feel that it is in its lowest-energy orientation, and no energy reduction can be obtained by rotating any dipole. Note that although the application of a weak field pointing along the z axis breaks the two-dimensional continuous symmetry of LT configurations, the rotation symmetry about the z axis is preserved, and one therefore still has a one-dimensional manifold of ground states, which is confirmed by our numerical simulations.

A somewhat similar result is obtained when the external field lies in one of the three main planes. In the limit of weak external field, it is energetically favourable for the dipoles to point perpendicular to the external field. However, no infinite degeneracy survives in this case. For example, if the external field lies in the xy plane (but not along x or y), the broken-symmetry configuration will have the dipoles pointing in the z direction. Choosing a configuration where the dipoles have a finite xy component and following the rules of constructing the LT dipole configuration would result in a situation where some dipoles are not perpendicular to the external field, which is not as energetically favourable as having all the dipoles perpendicular to the direction of the external field.

The situation becomes more complicated when the external field has finite components in all three directions. One can see that this case is trickier than the above two by noting that if we take any given dipole and set it to any direction perpendicular to the external field, the LT rules determining the orientations of the other dipoles in the lattice dictate that some dipoles will point in the same general direction as the external field (even if not exactly parallel to it). In our simulations, we find that the dipoles point along one of the three crystal axes, specifically the one that makes the largest angle with the external field. For example if we take an external field direction defined by the spherical coordinates $(\theta,\phi)=(\pi/6,\pi/6)$, the field has the angles 64, 75 and 30 degrees with the x, y and z axes, respectively, and therefore the dipoles will point along the y axis.

\subsection{Strong external field}
\label{Sec:StrongField}

An infinitely strong external field would obviously dominate over the dipole-dipole interaction and polarize all the dipoles parallel to the external field. We therefore take as the strong-field state the configuration where the dipoles are ordered in a fully ferroelectric type of state, with the orientation vectors $(p_x, p_y, p_z)$ being uniform for all the dipoles. In this state, as shown in the Appendix, the internal field induced by all neighbouring dipoles at the location of a central dipole vanishes, independently of the orientation direction of the dipoles. This result means that the internal field does not favour any rotation of any dipole away from the external field direction and confirms that for sufficiently strong external fields all the dipoles will align with the external field.

Note that the vanishing of the internal field in the ferroelectric state means that the ferroelectric configuration with axis parallel to the external field must be a dynamically stable state for any value of the field. The reason is that in this case every dipole is parallel to the total field at the dipole's location, which is a dynamically stable state. However, for fields smaller than a certain value (which turns out to be $5.589 U_0$ as discussed above) other configurations become energetically more favourable than the ferroelectric configuration. This situation can also lead to a hysteresis effect, as discussed e.g.~in Ref.~\cite{AbuLabdeh}.

\subsection{Intermediate field values}
\label{Sec:IntermediateField}

In the case where the external field lies in one of the main planes, the dipoles start off pointing perpendicular to the external field in the limit of very weak field. As the field strength increases, all the dipoles rotate by the same angle to become partially aligned with the external field. The angle is determined by the linear increase in polarization as the field is increased from zero to $5.589 U_0$. For field values above $5.589 U_0$, the dipoles are all aligned with the external field. For all values of the external field strength, the two-site translation symmetry is clearly preserved in all three directions \cite{Luttinger}. In other words, it suffices to perform calculations on a $2 \times 2 \times 2$ lattice to determine the dipole configuration in the bulk of the material.

For general field directions, the situation is more complicated because not all the dipoles make the same angle with the external field. In this case, we find that simulations on large lattices, e.g. $10 \times 10 \times 10$, give different results from simulations on a $2 \times 2 \times 2$ lattice. The $10 \times 10 \times 10$ configurations therefore do not have the two-site translation symmetry of the LT configurations. Otherwise the results calculated using the $10 \times 10 \times 10$ lattice and those calculated using a $2 \times 2 \times 2$ lattice would coincide with each other.

\begin{figure}[h]
\includegraphics[width=16.0cm]{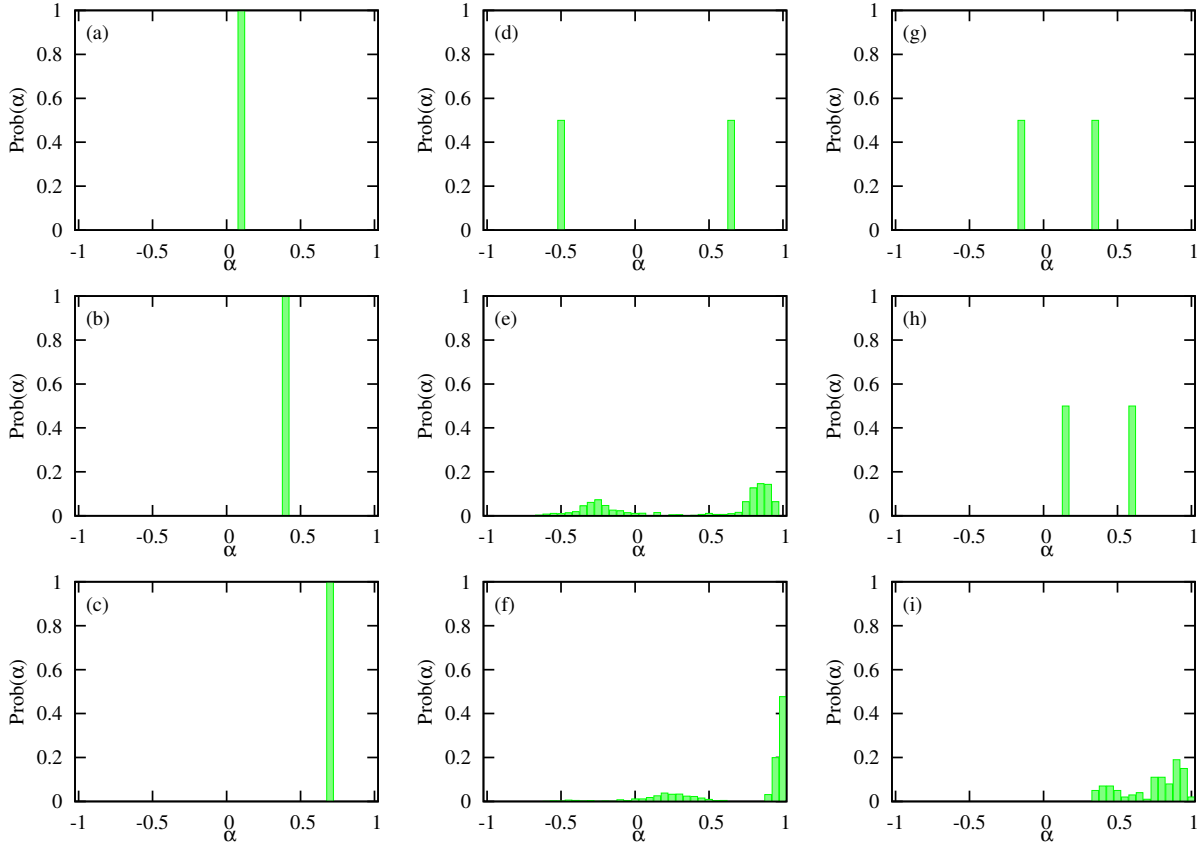}
\caption{Histograms showing the distribution of values for the individual dipole components $\alpha$ along the direction of the external field. From left to right the field direction is varied: $(\theta,\phi)=(\pi/2,\pi/8)$ (left column), the direction (1,1,1) (middle column) and $(\theta,\phi)=(\pi/6,\pi/6)$ (right column). From top to bottom the field strength is varied: $E_{\rm ext}=0.1 E_0$ (top row), $0.4 E_0$ (middle row) and $0.7 E_0$ bottom row.}
\label{Fig:PolarizationHistogram}
\end{figure}

Figure \ref{Fig:PolarizationHistogram} shows a histogram of the dipole components along the direction of the external field. Here we define the polarization of a single dipole along the field direction as $\alpha_i=\hat{p}_i\cdot\vec{E}_{\rm ext}/\left|E_{\rm ext}\right|$. When the field lies in one of the main planes, all the dipoles make the same angle with the field, and the histogram therefore contains only one peak that moves from $\alpha=0$ to $\alpha=1$ as the field strength is increased from $E_{\rm ext}=0$ to $E_{\rm ext}=E_0$. When the field lies outside the main planes, the histogram starts off with two peaks at $E_{\rm ext}=0$, as would be expected for a LT configuration in which the dipoles point along one of the main axes. When the field strength is increased and the two-site translation symmetry is broken, the histogram in general contains a large number of $\alpha$ values, which reflects the fact that one no longer has at most eight dipole orientations that are repeated throughout the lattice. The fact that the histogram turns into a broad distribution at intermediate values of the external field is also interesting because it demonstrates that the dipole configurations become complex, with no simple obvious pattern, even in the ground state. It is expected that such complex patterns also arise for larger lattices, such that the ground state would look disordered at intermediate field values.

\begin{figure}[h]
\includegraphics[width=16.0cm]{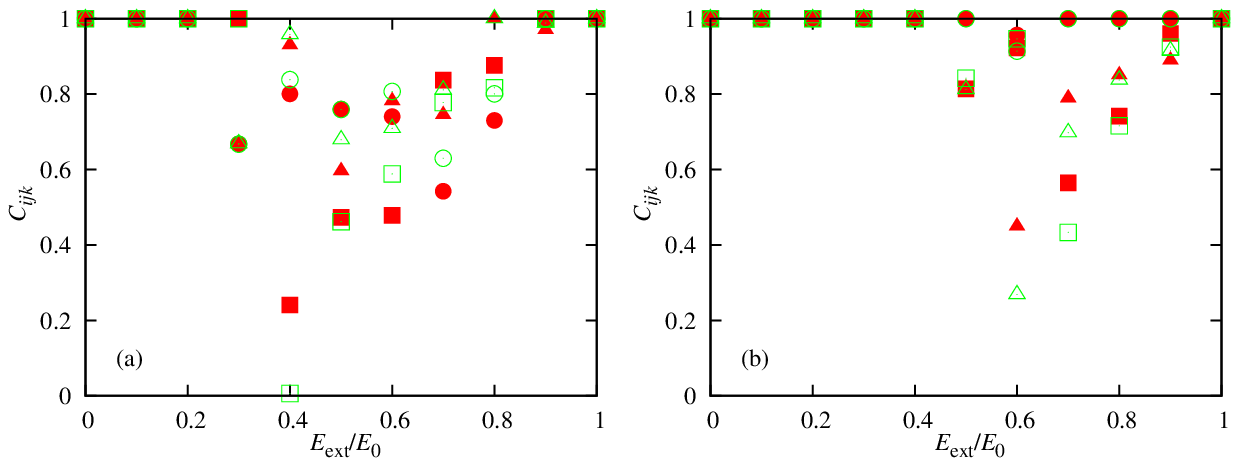}
\caption{The correlation function $C_{ijk}$ for different values of $(i,j,k)$: (2,0,0) [filled red squares], (0,2,0) [filled red circles], (0,0,2) [filled red triangles], (4,0,0) [open green squares], (0,4,0) [open green circles] and (0,0,4) [open green triangles]. Two field directions are used in the calculations: the direction (1,1,1) [Panel a] and $(\theta,\phi)=(\pi/6,\pi/6)$ [Panel b]. In the case of the direction (1,1,1), there is a rotation symmetry in which the three main axes are equivalent to each other. This symmetry is clearly not present in the dipole configuration, since for example the red symbols do not all coincide with each other.}
\label{Fig:CorrelationFunction}
\end{figure}

Figure \ref{Fig:CorrelationFunction} shows the dipole-dipole correlation function
\begin{equation}
C_{ijk} = \langle \hat{p}_{abc} \cdot \hat{p}_{a+i,b+j,c+k} \rangle,
\label{Eq:CorrelationFunction}
\end{equation}
where the average is taken over all the dipoles in the lattice. Interestingly, for the cases shown in the figure, the correlation function remains at its maximum value $C_{ijk}=1$ (meaning that the two-site translation symmetry is preserved) up to finite fraction of $E_0$ (specifically $E_{\rm ext}=0.2 E_0$ and $E_{\rm ext}=0.4 E_0$ for the two cases plotted in Fig.~\ref{Fig:CorrelationFunction}). This result indicates that a LT-like state persists for a finite range of external field values. At somewhat higher values of the field, the correlation function drops below one. The initial drop is rather quick, such that the dipole configuration changes suddenly from a perfectly ordered state to a highly disordered state. As the external field becomes stronger, the dipoles become increasingly aligned with the external field until they are fully polarized along the direction of the external field, which occurs when the external field $E_{\rm ext}$ reaches $E_0$. At this point $C_{ijk}$ returns to its maximum value $C_{ijk}=1$, because all the dipoles are aligned with each other by virtue of their alignment with the external field. The dipole-dipole correlations in this case have nothing to do with internal interactions, since a strong field would lead to the same correlations regardless of any details related to the internal interactions. Note that these somewhat artificial correlations can be suppressed in the theoretical analysis by defining the correlation function as $\langle \hat{p}_{abc} \cdot \hat{p}_{a+i,b+j,c+k} \rangle - P^2$. However, since our purpose of analyzing the correlation function is to investigate the periodicity in the dipole configuration, we use the definition given in Eq.~(\ref{Eq:CorrelationFunction}).

\section{Conclusion}
\label{Sec:Conclusion}

We have analyzed the behaviour of the ground state of a simple cubic dipole lattice with dipole-dipole internal interactions and interacting with an external field. This system resembles the high temperature phase of  methylammonium lead iodide, although it neglects the interaction of the dipolar cations with the inorganic lattice. In the special case where the external field lies in one of the main lattice planes, our results agree with the known result that the dipole configurations remain rather simple and give an exactly linear relation between the polarization and the external field. When the external field does not lie in any of the main lattice planes, we find a number of interesting results. The ground state becomes a complex dipole configuration with no simple regular structure at intermediate values of the external field, even though the polarization deviates only slightly from the linear dependence on the strength of the external field. This result is unchanged when we use somewhat different sample sizes, up to $20 \times 20 \times 20$ in some test simulations, suggesting that these results are not consequences of finite size effects. We also find that the perfect LT ordering persists for relatively weak external fields until we reach a certain value of the external field at which the system suddenly turns to a disordered state, before returning to an ordered state at strong fields. Another interesting result that we have found is the fact that the value of the external field at which the system becomes fully polarized coincides with the value of the internally induced field in the absence of any external field. Our results help elucidate the microscopic physics of dielectric and magnetic materials in external fields, including hybrid organic-inorganic perovskites, and can find applications in recently emerging artificial magnetic materials and superstructures.

\section*{Acknowledgment}

We would like to thank S.~Rashkeev for useful discussions. This work was made possible by NPRP grant \# 8-086-1-017 from the Qatar National Research Fund (a member of Qatar Foundation). The findings achieved herein are solely the responsibility of the authors.

\section*{Appendix: Internal field in the ferromagnetic state}
\label{Sec:Appendix}

\setcounter{figure}{0}
\setcounter{equation}{0}
\renewcommand\theequation{A\arabic{equation}}
\renewcommand\thefigure{A\arabic{figure}}

Here we show that the neighbouring-dipole-induced field vanishes in the ferromagnetic state, independently of the dipole orientation.

In the ferromagnetic state, all dipoles have the same components $(p_x, p_y, p_z)$. Because of the cubic symmetry, for every dipole at location $(x, y, z)$ there are dipoles at locations given by the permutations [e.g.~$(z, y, x)$] as well as mirror image locations [e.g.~$(x,-y, z)$]. Taking the dipole at the origin and the dipole at $(x, y, z)$, the first term in the dipole-dipole interaction energy (Eq.~\ref{Eq:DipoleDipoleInteractionEnergy}) has the factor
\begin{equation}
\vec{p}_i\cdot\vec{p}_j = p_x^2 + p_y^2 + p_z^2 = p^2.
\end{equation}
For the second term in Eq.~(\ref{Eq:DipoleDipoleInteractionEnergy}), taking the dipoles at the origin and at $(x, y, z)$, we find
\begin{eqnarray}
\left(\vec{p}_i\cdot\hat{r}_{ij}\right) \left(\vec{p}_j\cdot\hat{r}_{ij}\right) & = & \frac{\left( x p_x + y p_y + z p_z \right)^2}{r_{ij}^2} \nonumber \\
& = & \frac{1}{r_{ij}^2} \left( x^2 p_x^2 + y^2 p_y^2 + z^2 p_z^2 + 2 x y p_x p_y + 2 x z p_x p_z + 2 y z p_y p_z \right)
\end{eqnarray}
The cross terms, i.e.~the last three terms inside the brackets, can be ignored because for every nonzero coordinate, e.g.~$x$, there will be a partner term with the coordinate $-x$ giving the same term with the opposite sign. Each one of the first three terms has the square of a dipole Cartesian component multiplied by the square of one Cartesian component of the relative position vector. When combined with similar terms for permutations of $(x, y, z)$, we obtain products such as $y^2p_x^2$ and $z^2p_x^2$, such that the sum over all the permutations results in the value $\left(p_x^2 + p_y^2 + p_z^2\right)$ multiplied by the total number of permutations of $(x, y, z)$ and divided by 3. When we take into consideration the factor of 3 in the second term in Eq.~(\ref{Eq:DipoleDipoleInteractionEnergy}), we find that the two terms in the sum exactly cancel each other, independently of the orientations of the dipoles. We therefore find that the total energy and hence the local field at any dipole must vanish regardless of the orientation of the dipoles.


\begin{thebibliography}{99}

\bibitem{SpinModelBooks} See e.g. R. J. Baxter, {\it Exactly Solved Models in Statistical Mechanics} (Academic Press, London, 1982); H. E. Stanley, {\it Introduction to Phase Transitions and Critical Phenomena} (Oxford University Press, New York, 1987); S. Sachdev, {\it Quantum Phase Transitions} (Cambridge University Press, Cambridge, 2011).

\bibitem{Amico} L. Amico, R. Fazio, A. Osterloh, and V. Vedral, Rev. Mod. Phys. {\bf 80}, 517 (2008).

\bibitem{Eisert} J. Eisert, M. Cramer, and M. B. Plenio, Rev. Mod. Phys. {\bf 82}, 277 (2010).

\bibitem{Baikie2013aa} T. Baikie,Y. Fang, J. M. Kadro, M. Schreyer, F. Wei, S. G. Mhaisalkar, M. Graetzeld, and T. J. White, J. Mater. Chem. A {\bf 1}, 5628 (2013).

\bibitem{Stoumpos2013aa} C. C. Stoumpos, C. D. Malliakas, and M. G. Kanatzidis, Inorg. Chem. {\bf 52}, 9019 (2013).

\bibitem{Ashhab} S. Ashhab, M. Carignano, and M. E. Madjet, J. Appl. Phys. {\bf 125}, 163103 (2019).

\bibitem{Frost2014ab} J. M. Frost, K. T. Butler, and A. Walsh, APL Mater. {\bf 2}, 081506 (2014).

\bibitem{Frost2014aa} J. M. Frost, K. T. Butler, F. Brivio, C. H. Hendon, M. van Schilfgaarde, and A. Walsh, Nano Lett. {\bf 14}, 2584 (2014).

\bibitem{Leguy} A. M. A. Leguy, J. M. Frost, A. P. McMahon, V. Garcia Sakai, W. Kockelmann, C. Law, X. Li, F. Foglia, A. Walsh, B. C. O`Regan, J. Nelson, J. T. Cabral, and  P. R. F. Barnes, Nature Commun. {\bf 6}, 7124 (2015).

\bibitem{Rashkeev} S. N. Rashkeev, F. El-Mellouhi, S. Kais, and F. H. Alharbi, Sci. Rep. {\bf 5}, 11467 (2015).

\bibitem{Lahnsteiner} J. Lahnsteiner, R. Jinnouchi, and  M. Bokdam,  arXiv:1905.12540 (2019).

\bibitem{Bramwell} S. T. Bramwell, Nature {\bf 397}, 212 (1999).

\bibitem{Cioslowski} J. Cioslowski and A. Nanayakkara, Phys. Rev. Lett. {\bf 69}, 2871 (1992).

\bibitem{Aoyagi} S. Aoyagi, N. Hoshino, T. Akutagawa, Y. Sado, R. Kitaura, H. Shinohara, K. Sugimoto, R. Zhang, and Y. Murata, Chem. Commun. {\bf 50}, 524 (2014).

\bibitem{Gorshunov} B. P. Gorshunov {\it et al}., Nature Commun. {\bf 7}, 12842 (2016).

\bibitem{Skomski} R. Skomski, J. Phys.: Condens. Matter {\bf 15}, R841 (2003); D. J. Sellmyer and R. Skomski (Eds.), {\it Advanced Magnetic Nanostructures} (Springer, New York, 2006).


\bibitem{Honda} K. Honda and J. Okubo, Phys. Rev. {\bf 10}, 705 (1917).

\bibitem{Ornstein} L. S. Ornstein and F. Zernike, KNAW Proceedings {\bf 21}, 911 (1919).

\bibitem{Luttinger} J. M. Luttinger and L. Tisza, Phys. Rev. {\bf 70}, 954 (1946); {\it ibid.} {\bf 72}, 257 (1947).

\bibitem{Lax} M. Lax, J. Chem. Phys. {\bf 20}, 1351 (1952).

\bibitem{Berlin} T. H. Berlin and J. S. Thomsen, J. Chem. Phys. {\bf 20}, 1368 (1952).

\bibitem{Nijboer} B. R. A. Nijboer and F. W. De Wette, Physica {\bf 24}, 422 (1958).

\bibitem{Horvat} A. Horvat, L. Pourovskii, M. Aichhorn, and J. Mravlje, Phys. Rev. B {\bf 95}, 205115 (2017).

\bibitem{Kukhtin} V. V. Kukhtin and O. V. Shramko, Phys. Lett. A {\bf 128}, 271 (1988).

\bibitem{AbuLabdeh} A. M. Abu-Labdeh, A. B. MacIsaac, J. P. Whitehead, K. De’Bell, and M. G. Cottam, Phys. Rev. B {\bf 73}, 094412 (2006).

\bibitem{Galkin} A. Yu. Galkin and B. A. Ivanov, JETP Letters {\bf 83}, 383 (2006).

\bibitem{Bondarenko} P. V. Bondarenko, A. Yu. Galkin, B. A. Ivanov, and C. E. Zaspel, Phys. Rev. B {\bf 81}, 224415 (2010).

\bibitem{Johnston} D. C. Johnston, Phys. Rev. B {\bf 93}, 014421 (2016).

\end{thebibliography}
\end{document}